\documentclass[%
 reprint,
superscriptaddress,
 amsmath,amssymb,
 aps,
longbibliography,
]{revtex4-2}

\usepackage{graphicx}
\usepackage{color}
\usepackage{dcolumn}
\usepackage{bm}
\usepackage[normalem]{ulem}
\usepackage{varioref}
\usepackage{hyperref}
\usepackage{cleveref}

\begin{document}

\preprint{APS/123-QED}

\title{Scaling silicon-based quantum computing using CMOS technology}

\author{M. F. Gonzalez-Zalba}
\email{fernando@quantummotion.tech}
\affiliation{Quantum Motion Technologies, Windsor House, Cornwall Road, Harrogate, HG1 2PW, United Kingdom}
\author{S. de Franceschi}
\affiliation{Univ. Grenoble Alpes and CEA, IRIG/PHELIQS, 38000 Grenoble, France}
\author{E. Charbon}
\affiliation{\'{E}cole Polytechnique F\'{e}d\'{e}rale de Lausanne, Case postale 526, CH-2002 Neuch\^{a}tel, Switzerland}
\author{T. Meunier}
\affiliation{CNRS, Grenoble INP, Institut Neel, University of Grenoble Alpes, 38000 Grenoble, France}
\author{M. Vinet}
\affiliation{CEA/LETI-MINATEC, CEA-Grenoble, 38000 Grenoble, France}
\author{A. S. Dzurak}
\affiliation
{School of Electrical Engineering and Telecommunications, The University of New South Wales, Sydney, NSW, 2052, Australia}

\date{\today}

\begin{abstract}

As quantum processors grow in complexity, attention is moving to the scaling prospects of the entire quantum computing system, including the classical support hardware. Recent results in high-fidelity control of individual spins in silicon combined with demonstrations that these qubits can be manufactured in a similar fashion to field-effect transistors, create an opportunity to leverage the know-how of the complementary metal–oxide–semiconductor (CMOS) industry to address the scaling challenge at a system level. Here, we review the prospects of scaling silicon-based quantum computing using CMOS technology. We consider the concept of a quantum computing system, which we decompose into three distinct layers --- the quantum layer, the quantum–classical interface and the classical layer --- and explore the challenges involved in their development, as well their assembly into an architecture. Silicon offers the enticing possibility that all layers can, in principle, be manufactured using CMOS technology, creating an opportunity to move from distributed quantum-classical systems to integrated quantum computing solutions.

\end{abstract}

\pacs{Valid PACS appear here}
															
\maketitle

Quantum computers have demonstrated computational advantage~\cite{Arute2019, Zhong2020}.  As the technology continues to develop, scaling quantum computing systems as a whole becomes increasingly important and will be essential to building a machine with sufficient error-free computing resources to run quantum algorithms that can solve problems of societal value. Fault-tolerant quantum computing requires resilience against errors. Topological quantum computing models based on non-Abelian anyons -- such as Majorana zero modes -- offer protection at the qubit level~\cite{Sarma2015, Karzig2017}. However, the most technologically promising routes to fault-tolerant quantum computing are based on the standard quantum computing paradigms that use noisy qubits in combination with quantum error correction (QEC)~\cite{Lidar2013, Devitt2013, Campbell2017}. In this scheme, topological protection is achieved by distributing the logical information over a number of physical qubits, as long as each satisfy a maximum error rate in the combined initialization, manipulation and readout. The most forgiving QEC method, the surface code, sets a ~1\% upper bound~\cite{Fowler2012}. The exact physical qubit overhead (per logical qubit) depends strongly on the error rate but considering state-of-the-art qubit fidelities, it will likely be a figure in excess of a thousand. QEC is then expected to take the number of required physical qubits to many thousands and possibly millions for economically significant algorithms~\cite{Bauer2016,Reiher2017} and to many millions or billions for some of the more demanding quantum computing applications such as Shor's factorization algorithm~\cite{Shor1997}. Large-scale integration is hence a requirement to implement QEC schemes and a technological challenge for the most advanced quantum computing platforms relying on superconductors, semiconductors, ion traps or photonic circuits as the physical hosts for the qubits~\cite{Monroe2013, Devoret2013, Awschalom2013, Knill2001}. 

Developments in the field of nanodevice engineering have shown that qubits can be manufactured in a similar fashion to field-effect transistors (FET)~\cite{Maurand2016, Zwerver2021, Camenzind2021}, creating an opportunity to leverage the integration capabilities of the semiconductor industry to address the up-scaling challenge. Silicon-based quantum computing offers a number of key technological advantages favouring large-scale integration: (i) A small qubit footprint, of the order of $100\times100$~nm$^2$, and potentially commensurate I/O electronics~\cite{Veldhorst2017}; and (ii) compatibility with established very large-scale integration (VLSI) techniques of the CMOS industry that routinely manufacture billions of quasi-identical transistors on the size of a fingertip. These ingredients could allow compact fault-tolerant quantum processors that can fit in conventional cryostats without requiring the technically challenging development of new large-scale infrastructures~\cite{Lekitsch2017}. Furthermore, a quantum processing unit (QPU) will likely be a sub-module of a larger information processing system that also contains analog and digital electronics. In such a scenario, CMOS could enable hybrid integration of quantum and classical technologies, facilitating data management and fast information feedback between them. And, even before a fault-tolerant quantum computer is built, compact CMOS manufacturing could deliver multi-core NISQ QPUs~\cite{Cai2020}, ideal for hybrid quantum-classical algorithms that benefit from massive parallelisation~\cite{Peruzzo2014,farhi2014}.

\begin{figure*}[htbp]
	\centering
		\includegraphics[width=1.00\textwidth]{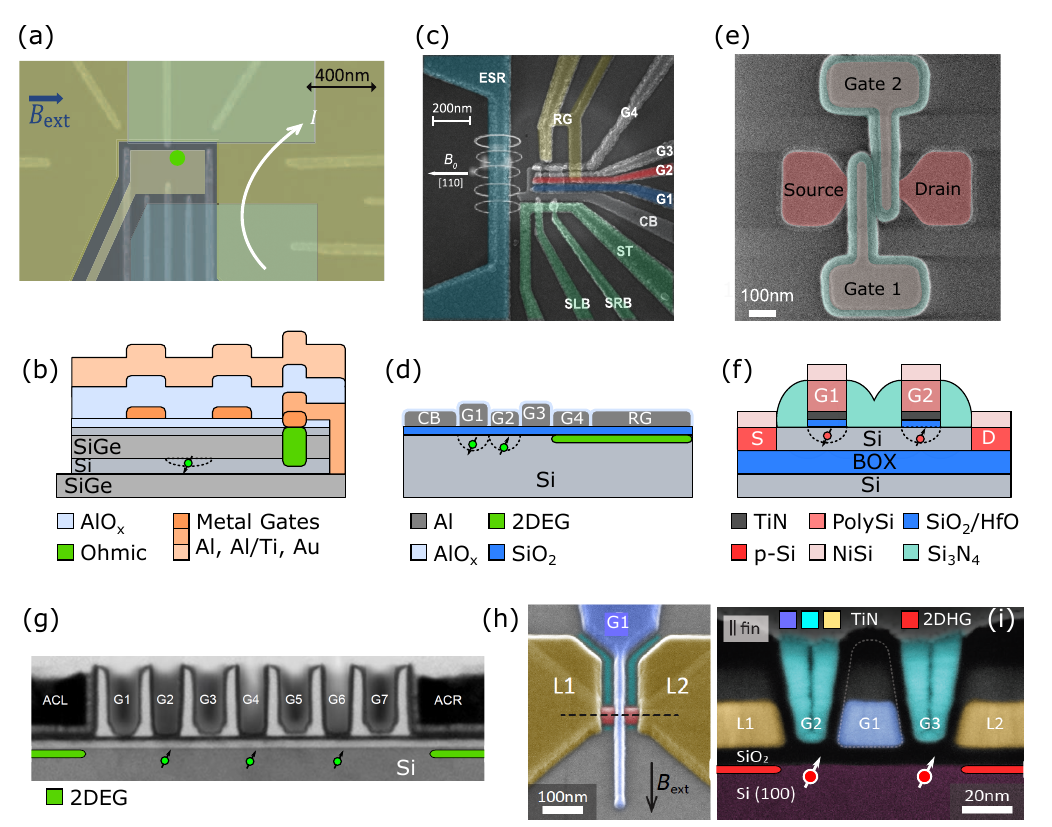}
	\caption{Silicon QD devices. (a) Scanning electron microscope (SEM) image of an accumulation mode Si/SiGe heterostructure. Two layers of gates, bottom (light grey) and top (dark green) are designed to form two QDs (centre of image) and a single-electron transistor for readout (right). The structure contains a micromagnet in an upper metal layer (light green) to produce a magnetic field gradient. The green dot indicates the position of a QD used in ref.~\cite{KawakamiE.2014}. (b) Cross-sectional schematic of a Si/SiGe QD device. (c) SEM image of a metal-oxide-semiconductor multi QD device with QD gates (G1-4), confinement gate (CB), reservoir gate (RG), an integrated single-electron transistor (green) and microwave antenna for magnetic resonance spin control (blue) used in ref.~\cite{Huang2019}. (d) Cross-sectional view of the MOS QD device in (b). (e) SEM image of a CMOS p-type double QD on an etched silicon-on-insulator nanowire used in ref.~\cite{Maurand2016} and schematic cross section (f). SEM (g) and TEM image (h) of a hole spin double QD device in a self-aligned double layer gate structure. L1(2) are 2 dimensional hole gas (2DHG) accumulation layers. G2, 3 are qubit control gates and G1 control the tunnel coupling~\cite{Camenzind2021}.}
	\label{Fig1}
\end{figure*}

In addition to the technological benefits, silicon offers favourable physical properties that allow qubits with long coherence times to be created. The qubits are encoded by localized spins at deep cryogenic temperatures and finite magnetic fields. The simplest example is the spin $\frac{1}{2}$ of a single electron (or hole) electrostatically confined in a quantum dot (QD)~\cite{Loss1998}. Alternatively, the nuclear spin of individual dopants can be used~\cite{Kane1998}. Quantum dots can now be manufactured on demand in a silicon metal-oxide-semiconductor (MOS) nanodevice (typically Metal-SiO$_2$-Si)~\cite{Veldhorst2014} or in a Si/SiGe heterostructure~\cite{KawakamiE.2014} (Fig.~\ref{Fig1}a-d). Recently, large enhancements in spin coherence have been achieved by isotopic enrichment of the silicon lattice. The 5\% naturally occurring spin-carrying isotope, $^{29}$Si, is the major source of decoherence in silicon, whereas the dominant $^{28}$Si isotope has zero nuclear spin. By enriching to nearly pure $^{28}$Si, with only 800 ppm of $^{29}$Si remaining, inhomogeneous spin-dephasing times ($T_2^*$) and spin-coherence times ($T_2$) exceeding 100~$\mu$s and 20~ms, respectively, have been measured for electron spins, placing silicon as one of the most coherent solid-state systems in nature~\cite{Veldhorst2014}. For optimal performance, silicon qubits are cooled down to a few tens of millikelvin under magnetic fields of the order of 1~T but these parameters may be relaxed in the future to allow operation above 1~K~\cite{Yang2020a, Petit2020} and at just 150~mT~\cite{Zhao2019}.

Silicon spin qubits are initialized and readout using spin-to-charge conversion techniques~\cite{Elzerman2004, Johnson2005, Morello2010, Maune2012, Fogarty2018}, they are coherently manipulated via magnetically- or electrically-driven electron-spin-resonance for single-qubit operations~\cite{PioroLadriere2008, Pla2012, KawakamiE.2014} and spin-exchange-based methods for two-qubit logic~\cite{Veldhorst2015}. Thanks to the increase in coherence, advancements in high-frequency readout techniques, and optimized spin projection mechanisms, all these steps have now been performed with fidelities above the requirements of the surface code. More particularly, a gate set at the surface code error threshold has been recently demonstrated on a two-qubit device~\cite{Xue2021a, Noiri2021}. These are promising initial results for this relatively recent approach to quantum computing, and indicate that building a fault-tolerant quantum computer based on silicon technology is a realistic proposition. However, several technological challenges lie ahead. So far, most advances have been achieved with small-scale devices (one-, two- and recently six-qubit systems) fabricated in academic cleanrooms, offering relatively modest level of process control and reproducibility~\cite{Zajac2017, Watson2018, Huang2019, Vandersypen2021}. From this point onwards, a route to increase fidelities well beyond fault-tolerant thresholds in a reproducible way across large arrays of qubits needs to be established. Recent results on qubits fabricated in industrial settings~\cite{Maurand2016, Zwerver2021, Camenzind2021} may shed some light on how to achieve this goal, see Fig.~\ref{Fig1}e-h.

In this Review, we explore the scaling prospects of quantum computing systems based on CMOS technology. We first consider the concept of a quantum computing system and its different elements. We then examine the challenges at the quantum layer and the challenges involved when designing classical CMOS circuits at cryogenic temperatures, which affect both the quantum–classical interface and the classical layer. Finally, we consider the functional assembly of the layers into a system architecture.

\section*{A Quantum Computing System}\label{QPU}

\begin{figure*}[htbp]
	\centering
		\includegraphics[width=1.0\textwidth]{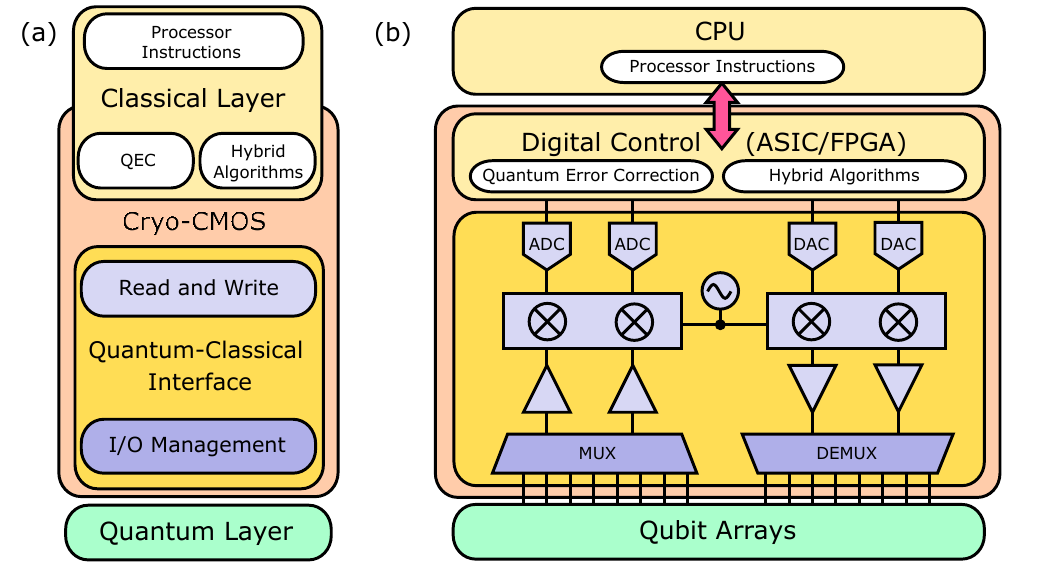}
	\caption{A quantum computing system. (a) Schematic representation of the main layers of a QCS. (b) A more detailed representation of a QCS including the interconnection between modules. The quantum-classical interface uses (de)multiplexers to facilitate input/output data management. These components could operate using time-domain and frequency-domain multiplexing. Control and readout micro- and radio-wave tones are produced by IQ modulation and amplification and readout signals are detected after amplification via IQ demodulation. The digital controller is used for fast feedback between the classical and quantum units, especially for hybrid quantum-classical algorithms and quantum error correction, and also to send the quantum computer instructions received from a classical computer. It interfaces with the rest of the layers via ADC/DACs.}
	\label{Fig2}
\end{figure*}

The problem of scaling requires a shift in the thinking process beyond the few-qubit quantum processor proof-of-principle demonstrations to a full stack perspective: that is, a quantum computing system (QCS)~\cite{Jones2012}. Researchers have already started to tackle this problem, producing blueprints of what a large-scale quantum computer could look like in silicon~\cite{Veldhorst2017, Vandersypen2017, Li2018, Boter2019}. These proposals share a common basic idea. In particular, the future silicon QCS should be composed of three distinct layers -- the quantum layer or quantum processing unit (QPU), the quantum-classical interface and the classical layer -- all of which could potentially be manufactured using CMOS technology, at least to a certain extent (Fig~\ref{Fig2}a,b).

\begin{figure*}[htbp]
	\centering
		\includegraphics[width=1.00\textwidth]{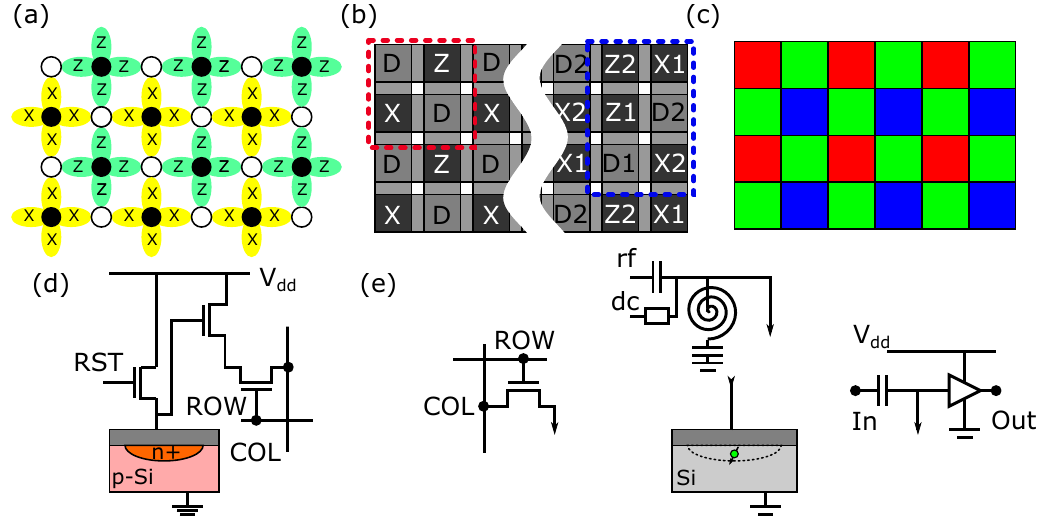}
	\caption{Two-dimensional arrays. (a) Graphical representation of a two-dimensional qubit array with nearest neighbour interactions necessary to implement the surface code. The hollow circles are data qubits and the black circles are error detection qubits, either X or Z syndrome, redrawn from~\cite{Fowler2012}. Yellow and green ellipses represent two qubit interactions with the X- and Z- syndrome qubits, respectively. (b) Schematic representation of two ways of assigning data (grey) and syndrome qubits (dark grey) in a two-dimensional gate electrode array with tunable exchange gates (light grey). (Left) Standard 2x2 qubit sublattice (delimited by the red dashed line) requiring vertically integrated readout~\cite{Vinet2018}. (Right) Extended 2x3 qubit sublattice (delimited by the blue dashed line) with two additional syndrome qubits to enable readout in the plane~\cite{Veldhorst2017}. (c) Graphical representation of a digital image sensor formed by a two-dimensional array of photodiodes (pixels) combined with filters to detect specific wavelengths (RGB). (d) Schematic of a three-transistor active pixel sensor, including a reset transistor for the photodiode (RST), a source-follower readout transistor and a selection transistor addressed via row-column inputs. (e) Cross-section schematic of a single-electron MOS quantum dot with possible combinations of classical electronics. Left) Row-column access transistor for 2D addressing. Top) $LC$ resonant circuit for dispersive readout where the MOS structure plays the role of variable capacitor. Right) Local amplification at the cell level.}
	\label{Fig3}
\end{figure*}

The configuration of the quantum layer is primarily dictated by the physical requirements of the surface code (Fig~\ref{Fig3}a): a two-dimensional distribution of physical qubits with nearest-neighbour interactions~\cite{Fowler2012}. The qubits are split into data qubits (qubits in which the computational quantum states are stored) and measurement qubits (qubits performing error detection). Measurement qubits come in two types: X and Z syndrome qubits, which contribute to the measurement of bit-flip and phase-flip errors, respectively. Each X-type and Z-type qubit is made to interact in an anticlockwise sequence with the four adjacent data qubits and then measured to extract information of the system (stabilizer). The stabilizers can be measured repeatedly without perturbing the quantum state of the system. However, once one or more errors occur, the outcome of a stabilizer measurement changes. By correlating stabilizer outputs, the location of an error can be identified and then corrected. In silicon, the surface code could be implemented in a $^{28}$Si substrate, by distributing, in a square array, individually-addressable nanometre-scale MOS structures to form few-electron QDs, plus additional gates in between to control, locally, the exchange interaction between spins. Given the projective nature of spin readout, the surface code could be implemented in a $2\times3$ QD sublattice, using planar designs~\cite{Veldhorst2017}, or in a canonical $2\times2$ sublattice, although this would involve using 3D integrated structures (Fig~\ref{Fig3}b)~\cite{Vinet2018}. 

The quantum-classical interface will handle the control and readout of the individual qubits, as well as the routing and input/output data management of the large number of signals required to operate the QPU~\cite{Reilly2015}. This layer contains signal generators, (de)modulators, analog-to-digital (ADCs) and digital-to-analog converters (DACs), amplifiers for control and readout and (de)multiplexing for I/O management (Fig.~\ref{Fig2}b).

The classical layer is designed to assist in the QEC process by taking the outputs of the QPU via the quantum-classical interface and correlating them, in classical logic, with a concrete error type, location and time step. Errors can be generally corrected in software but in order to enable universal quantum computation, including non-Clifford operations, fast feedback between readout and control would be necessary. Furthermore, the classical layer is expected to compile the quantum algorithms into a sequence of interleaved control and readout steps of the relevant qubits. This layer contains a processing unit (field-programmable gate array (FPGA); or application-specific integrated circuit (ASIC)) that, within the cooling power constraints of cryogenic systems, should be placed in close proximity to the rest of the layers to reduce latency in feedback operations.

The physical architecture of the QCS is still a matter of debate. In fact, it is likely that a number of approaches could be followed. The flexibility in the physical arrangements of the layers and the square-grid distribution of the QDs in the QPU evoke image sensors as an example of the foreseeable future for CMOS-based quantum computing (Fig~\ref{Fig3}c). Charge coupled devices (CCDs) or CMOS image sensors, competing image technologies, both fulfill their image recording purposes but with a different set of specifications given their different levels of integration~\cite{Cressler2009}. In both technologies, the physical layout is constituted by a two-dimensional array of unit cells (pixels). In the case of CCD, charge-to-voltage conversion occurs via shifting charges sequentially to a global amplifier sitting at the periphery. However, in the case of CMOS sensors, such as active-pixel sensors, readout occurs in a distributed manner with a first stage of amplification at the pixel level, followed by column and global amplification at the periphery (Fig~\ref{Fig3}d). The different architectures result in a different set of technical specifications that can be tailored to the specific application. CMOS-based quantum computing could follow a similar path of development in which some architectures could have either local or global electronics for control and readout. Global techniques are likely to simplify the integration process and deliver architectures sooner, while local integration will provide enhanced functionality. We note that although the analogy works in first order approximation, qubits present radical differences to pixels. Qubits have comparably smaller footprints than photodiodes and qubit-to-qubit interactions need to be managed. These differences impose restrictions for local electronics that will require technological innovation beyond state-of-the-art image sensors. 

\section*{\label{SoA} Challenges at the quantum layer}

There is flexibility in the path ahead as the optimal qubit cell is still to be defined (Fig~\ref{Fig3}e). However, the abstract description of a quantum machine presented above has some clear consequences on the required conditions to build a QCS. We thus highlight areas where silicon quantum information processing could benefit from the know-how of the CMOS industry starting with the QPU.

\subsection*{Qubit arrays}

The first challenge involves to the use of commercial-scale CMOS manufacturing lines to enable the fabrication of dense two-dimensional arrays of individually addressable gate-defined QDs with gate-controlled tunnel barriers. The unit cell of the qubit is likely to contain additional commensurable electronics either in-plane or in a 3D geometry. Here, we exclusively concentrate on the QD array. The characteristic footprint of the gate electrodes that have been used to defined few-electron QDs in CMOS platforms has been of the order of $40\times40$~nm (width and length) with a gate pitch of 70~nm ~\cite{Maurand2016, Corna2018, Crippa2018, ansaloni2020}, although these dimensions may differ depending on the exact gate stack. The dimensions are in line with the 22~nm CMOS node and are not likely to pose a critical problem. However, the routing of the individual lines is a challenge. To allow 2D geometries, ideally, the gate interconnect should be placed directly above the location of the QDs, a feature that is yet not possible in modern technology nodes. In addition, the necessity to have a tightly packed exchange gate in between adjacent QD gates to control the interaction between spins, presents a challenge. Currently, there is no standard CMOS technology node that could deliver the required exchange gate layout. Research efforts are being devoted to developing multi-gate-layer processes in CMOS foundries that will enable configuring exchange gates. So far, two technological paths have been pursued to introduce the exchange gates: (i) three levels of overlapping gates~\cite{Veldhorst2015, Zajac2016} and (ii) self-aligned barrier gates~\cite{Geyer2021, Zwerver2021}.

Currently, modules containing linear and bilinear arrays of QDs are being explored~\cite{Hutin2019, Chanrion2020, ansaloni2020,gilbert2020,duan2020} in which NISQ algorithms~\cite{Cai2020} or logical qubits could be implemented~\cite{Jones2018}. This modular approach could help to tackle the scaling problem in successive approximations, first by producing 1D arrays, then by combining them in 2D arrays with sparse connectivity. This concept would leave space between modules that could then be used to alleviate some of the problems associated with the gate contact routing and the location of the control and readout electronics. The idea could help to develop compact qubit unit cells with embedded electronics to finally produce a fully-connected 2D architecture. The options to build up complexity by coupling individual modules are varied. A possibility could be to directly connect linear or bilinear arrays of qubits at square junctions where couplings between x- and y-oriented arrays could be engineered. As we shall see later, it has also been suggested that medium range solutions like QD couplers~\cite{Malinowski2019} and floating gates~\cite{duan2020} or long range coherent links such as photon-mediated interactions between spins~\cite{Burkard2020} or even physical transfer of spins using shuttling~\cite{Yoneda2020} could be used to couple coherently distant modules.

\subsection*{High-fidelity control}

When using electrons as the spin carrying particle, silicon-based spin qubits coming from research laboratories have so far reached single-qubit fidelities exceeding 99.9\% in 200 ns~\cite{Yoneda2017}, and two-qubit fidelities of 98\% in 5 $\mu$s~\cite{Huang2019} and recently over 99\%~\cite{Xue2021a, Noiri2021}. While there is room for improvement, the objective is to achieve gate-set fidelities well above 99\% across a large array of qubits on timescales of the order of a microsecond or less to enable fast QEC protocols and minimise computation time. 

Single-qubit rotations are typically achieved with electron spin resonance (ESR) techniques that require the delivery of pulsed oscillatory magnetic or electric fields (as in the case of electric dipole spin resonance (EDSR)), in the 5 to 40~GHz range. Electron spin resonance requires either the fabrication of on-chip antennas~\cite{Dehollain2012} to provide localized control of groups of qubits (Fig.~\ref{Fig1}c), or the placement of the chip within a uniform oscillatory magnetic field, such as that produced by a resonant microwave cavity to enable control of the entire qubit array~\cite{Kane1998}. The resonant  microwave signal can be pulsed to drive single spin rotations in-phase (X rotations) or in quadrature (Y rotations)~\cite{Pla2012}. Alternatively, a continuous wave signal can be applied and qubits can be tuned in and out of resonance making use of the Stark shift using local pulses at the qubit gate~\cite{Laucht2015}. Stark shift can also be used to realise Z rotations by exploiting the electrical tunabillity of the g-factor~\cite{Camenzind2021}. X and Y rotation rates in experiments to date on electron spins have been limited to a few MHz due to the heat associated with producing B-field pulses with sufficient amplitude at the qubit location. 

While on-chip microwave transmission line antennas have been successfully used for ESR control of few-qubit systems~\cite{Veldhorst2014, Veldhorst2015, Huang2019}, the approach does not scale well as the number of antennas would grow with the number of qubits, posing complex microwave challenges and subjecting the processor to heat generated by the microwave currents passing through the transmission lines.

One solution is to use a global AC magnetic field~\cite{Kane1998}. The most obvious candidate for such a global ESR field is to use a conventional 3D microwave cavity. However, the metal gates, interconnects and bond wires adversely affect the quality factors of such resonators~\cite{Kong2015}, while the quantum processor chip is subject to alternating electric fields, generating induced currents that could interfere with the operation of the QPU and the control and readout electronics. In recent experiments, however, a new type of compact dielectric resonator, constructed from potassium tantalate (KTaO3), has been used to demonstrate ESR of single electron spins in a SiMOS QD device~\cite{Vahapoglu2020} and coherent control~\cite{Vahapoglu2021}. The KTaO3 is a quantum paraelectric material that has a high dielectric constant at cryogenic temperatures which serves to concentrate the AC magnetic field it generates into a compact microwave mode volume. This enables large, and uniform, AC magnetic fields to be delivered to the qubit plane, while minimising the external microwave power that needs to be delivered to the resonator.

In contrast with ESR control, EDSR induces spin rotations by making use of spin-orbit coupling that converts oscillatory electric fields into oscillatory magnetic fields in the reference frame of the electron~\cite{PioroLadriere2008, Kawakami2016}. Typically, micro-magnets are necessary to generate intense magnetic field gradients that translate into synthetic spin-orbit coupling with sufficient strength to drive coherent spin rotations. Using Co micro-magnets in split geometries in a plane above the qubit layer (Fig.~\ref{Fig1}a), rates up to tens of MHz have been reached~\cite{Yoneda2017}. However, the increased spin-orbit coupling created by the micro-magnet field gradient also makes the qubit susceptible to electrical noise, which can limit $T_2$~\cite{Yoneda2017, Struck2020} and may substantially reduce the relaxation time, $T_1$~\cite{Borjans2019}. So far the split magnet geometry has been appropriate to control and independently address one-dimensional arrays of qubits~\cite{Vandersypen2021, Simion2020}. For two-dimensional arrays, the micromagnet geometry will need to be redesigned to provide a suitable magnetic field gradient and addressability across the array, for example, by miniaturization to nanometric structures and by placing multiple magnets in grids~\cite{Singh2017}. The impact of these new geometries on the strength of the magnetic field gradients will need to be studied and, if insufficient, new materials with higher remanence will need to be explored.

On the other hand, intrinsic spin-orbit coupling, of which there is enhanced evidence in low symmetry QDs~\cite{Corna2018}, or in holes rather than electrons~\cite{Crippa2019}, may facilitate scaling since manipulation will not require additional elements in the qubit cell. In fact, very promising recent results have shown that finFET-based single hole spin qubits can be controlled close to the fault-tolerant threshold at 1.5~K, thanks to the fast spin-orbit-driven spin rotations (147~MHz) and the weak hyperfine coupling leading to $T_2^*=440$~ns~\cite{Camenzind2021, Bosco2021}.

Two-qubit gates are based on the exchange interaction between neighbouring spins. The exchange strength ($J$), which depends on the wavefunction overlap between participating spin particles, can be varied by changing the voltage detuning between QDs (asymmetric tuning)~\cite{Veldhorst2015} or by tuning the potential barrier between QDs (symmetric tuning)~\cite{Martins2016,Reed2016, Shim2018}. Symmetric tuning is less sensitive to charge noise allowing exchange coupling operations at a sweet spot where the $J$ has zero derivative with respect to voltage detuning. The short-range nature of the exchange interaction and the preference for symmetric tuning, impose the tightest restriction on the gate layout as explained above: a QD gate pitch of the order of 70~nm and an exchange gate in the space between of adequate footprint to tune the exchange interaction by several orders of magnitude between the ON and OFF states (when $J/h$ should be much smaller than the single- and two-qubit gate speeds). 

Two-qubit gates can be performed in two distinct ways, either via exchange modulation or resonantly. Exchange modulation involves pulsing the exchange interaction. Depending on the pulse scheme, a CPhase~\cite{Veldhorst2015} or a $\sqrt{\text{SWAP}}$ gate~\cite{Maune2012,He2019} can be implemented. In the latter case, a subnanosecond two-qubit gate has been demonstrated for P-donor spins in silicon~\cite{He2019}, emphasizing the speed advantages of gate-voltage pulses over resonant two-qubit gate schemes. Resonant two-qubit gates involve turning on the exchange interaction and performing ESR/EDSR on the coupled two-spin system to perform a CROT gate~\cite{Zajac2017,Huang2019}. CPhase gates set the benchmark for two-qubit gate fidelity: 99.5\% in 100~ns~\cite{Xue2021a}.

Besides two-qubit gates driven by direct exchange interaction, other proposals suggest incorporating larger multi-electron QDs that could extend the spin-spin interactions beyond nearest neighbours~\cite{Cai2020}. These QD mediators have been demonstrated to extend the spin interaction in an electrically-controlled manner over 800~nm in GaAs~\cite{Malinowski2019}, and could free up space in between qubits for gate routing while maintaining the full 2D connectivity. However, mediator QD level spacing must be larger than the energy of the thermal fluctuations and the excitation spectrum of the control voltage pulses~\cite{Malinowski2018}. Given the smaller effective mass of silicon compared to GaAs and the dependence of the level spacing on the size of the dot, this will reduce the range over which fast spin interactions could be driven in silicon by approximately a factor of 3. An experimental proof of mediator dots in silicon remains to be demonstrated. Another proposal that could extend the range of qubit interactions over intermediate distances is that of floating gates~\cite{Shulman2012}. However, due to its electrostatic nature, it would be most effective in qubits implemented in the magnetic number subspace $m_\text{s}=0$ of two-particle singlet-triplet qubits~\cite{Petta2005}, whose charge density is strongly dependent on the total spin number.

Going beyond one and two-qubit interactions, control at scale requires careful management of crosstalk between closely spaced qubits. Research should be directed towards assessing how manipulating target qubits impacts idling qubits (or even simultaneously addressed qubits) and determine compensation signals to minimise this crosstalk. Benchmarking of the optimal gate-set in terms of fidelity, operation timescales and impact on idling qubits must be undertaken.

\subsection*{\label{Readout}High-fidelity readout}

To implement fast feedback in active error correction protocols, high-fidelity readout must be performed in timescales significantly shorter than the spin coherence time~\cite{Laucht2021}. Ideally, this timescale should be shorter than the duration of single- and two-qubit gates to avoid becoming the bottleneck in QEC. The latter requires achieving readout fidelities well above 99\% in a few microseconds or less. Readout of spins in silicon is achieved via spin-dependent tunnelling processes such as \textit{Elzerman} readout~\cite{Elzerman2004} or Pauli spin blockade~\cite{Ono2002}, where spins tunnel selectively to a charge reservoir or to a spin polarized QD, respectively. Tunnelling under these conditions translates the spin information to a charge-based signal that can be read using sensitive electrometres.


The most commonly used are three-terminal charge sensors such as the single-electron transistor (SET)~\cite{Kastner1992} where a current flowing through the device is strongly dependent on the local charge environment. The SET can be used to detect single-electron tunnelling events in the time domain but its bandwidth is limited to a few tens of kilohertz by $RC$ parasitics at its output port. The bandwidth can be extended to hundreds of kilohertz using amplifiers in close proximity~\cite{Guevel2020} or even to tens of megahertz when using high-frequency $LC$ impedance matching techniques, i.e. the rf-SET~\cite{Schoelkopf1998, Angus2008}. With these two approaches, readout fidelities of 99.9\% in 6~$\mu$s~\cite{Curry2019}, and 99\% in 1.6~$\mu$s~\cite{Connors2020} have been achieved, respectively. These electrometres need to be placed in close proximity to the qubits and require two charge reservoirs to function complicating the use of this method at scale in dense qubit arrays.

More compact approaches are currently being developed. A two-terminal charge sensor, the rf single-electron box (SEB)~\cite{House2016, cirianotejel2021}, uses dispersive readout techniques to detect the variable capacitance of a QD. Changes in the surrounding charge environment modify the bias point of the SEB, which in turn produce a rf response conditional to the charge state of the sensed element. Although its demonstrated fidelity is still limited (99\% in 1~ms~\cite{Urdampilleta2019}), it may prove a valuable technique for upscaling. An even more compact method, \textit{in-situ} dispersive readout (IDR), simplifies the architecture by removing the charge sensor and charge reservoirs altogether and directly embedding the qubit in a radio- or microwave frequency electrical resonator~\cite{Wallraff2004, Colless2013, GonzalezZalba2015, Mi2018, Pakkiam2018, West2019, Crippa2019, Ibberson2021}. The state of the qubit, which in effect is a state-dependent capacitor~\cite{Mizuta2017}, can be directly inferred from the oscillatory state of the resonator. With this methodology a readout fidelity of 98\% in 6~$\mu$s has been obtained~\cite{Zheng2019}.

High-frequency techniques bring the benefit of simultaneous readout via frequency-domain multiple access (FDMA)~\cite{Hornibrook2014}. Additionally, time-division multiple access (TDMA) techniques could reduce the total number of resonators by performing qubit readout in a sequential manner using a common resonator~\cite{Schaal2019}. Furthermore, dispersive readout techniques, either rf-SEB and IDR, not being limited by shot-noise, can be used in conjunction with quantum-limited Josephson parameter amplification to speed up the readout~\cite{Schaal2020}.

In the case of high-frequency techniques, the footprint of the resonator (or inductor in lumped-element configurations) poses a significant challenge in terms of scaling with typical values in excess of $100\times 100$~$\mu$m~\cite{Zheng2019}. To solve this challenge, high inductance density materials will be necessary. Josephson metamaterials, formed by arrays of Josephson junctions, present one of the highest inductance per unit length and may be a compact solution~\cite{Stockklauser2017} but their integration within a CMOS process appears to be complex. High kinetic inductance materials~\cite{Samkharadze2016}, used for example in ultra sensitive detectors for astronomy~\cite{Mazin2004}, could be a better suited solution. Their inductance per unit length is increased over that of a single wire by a factor $\lambda_\textit{L}^2/t$, where $t$ is the thickness of the superconducting strip and $\lambda_\textit{L}$ is the London penetration depths. Research on industry-compatible high-kinetic inductance materials like TiN, with one of the highest reported London penetration depth (730~nm~\cite{Vissers2010}) and an inductance of $L_\text{K}=234$ pH/sq in 8.9~nm thin films~\cite{Shearrow2018}, could drastically reduce the resonator footprint to narrow strips a few micrometers long. Granular aluminium with $L_\text{K}=2$ nH/sq could be an even more compact alternative~\cite{Gruenhaupt2018, Gruenhaupt2019}.

If SET readout is to be pursued, research efforts should be directed at increasing the bandwidth even further by directly integrating low-power low-noise amplifiers on chip with minimal footprint capable of addressing several SETs via multiplexing techniques. The community should also think of ways to go beyond charge sensors and IDR by adapting concepts from classical electronics to resolve the fF-scale capacitance associated with quantum tunnelling between adjacent QDs~\cite{Esterli2019,Maman2020b}. A compact solution that could be integrated on-chip with a footprint commensurable to the qubit size will facilitate the massively parallel readout required for QEC codes. 

\subsection*{\label{Variability}Qubit variability}

One of the greatest challenges is the necessity to manufacture high-fidelity qubits at scale. Although VLSI technology guarantees a high level of reproducibility, variability acquires a much higher degree of importance in the quantum realm, since quantum device performance varies significantly with parameters like the tunnel coupling or the valley-splitting, both of which can be affected by a single atomic defect~\cite{Culcer2010,Rahman2011b,Ferdous2018}. Special emphasis should be put on the quality of the interfaces and on the purity and crystallinity of the materials.

From a channel material perspective, the development of isotopically enriched Si or Si/Ge stacks will be necessary to provide nuclear spin-free active substrates, which are key to suppressing the spin dephasing associated with the hyperfine interaction. Modules including $^{28}$Si enriched silane for Si, as well as $^{73}$Ge depleted germane for SiGe will need to be developed. Critical issues will need to be addressed such as the correct level of isotopic enrichment, the optimal thickness of the isotopically enriched channel taking into account performance, cost and the thermal budget to cope with the self-diffusion between natural and enriched silicon layers~\cite{Mazzocchi2019}. 

From a gate stack/dielectric perspective, material development remains critical since charged defects can affect the static and dynamic properties of the qubits. Stacks with minimized trapped charge densities below 10$^{11}$ cm$^{-2}$ will be required. This is likely to rule out high-k dielectrics known to have a large density of defects at the SiO$_2$/high-k interface (10$^{11}-10^{12}$ cm$^{-2}$) and put an emphasis on high quality Si-SiO$_2$ interfaces, which can have defect densities as low as $5\times 10^{10}$ cm$^{-2}$. Furthermore, due to the necessity to operate at cryogenic temperatures, the effect of thermal contraction of different materials will need to be studied and minimised. Thermal contraction mismatch between the gate stack and the device body can lead to defect/strain generation~\cite{Lo2015,Thorbeck2015} resulting in enhanced variability in the physical location of the QDs. Doped polycrystalline silicon gates are likely to minimise these effects. Tungsten, with a thermal expansion coefficient similar to that of silicon ($4.5\times 10^{-6}$~K$^{-1}$ vs $2.6\times 10^{-6}$~K$^{-1}$), may be an interesting candidate for lower resistivity gate material. The impact of metal gate granularity on the variability of the gate voltage for QD formation should also be minimised to avoid workfunction fluctuations~\cite{Zhang2014,Zeng2017,Brauns2018}.  

Considering now the variability in spin qubit operation frequencies, this is primarily determined by the variability in electron (or hole) G-tensor due to spin-orbit coupling, which in turn is influenced by a number of device and materials-related parameters, including interface roughness~\cite{Huang2017} and valley occupancy~\cite{Veldhorst2015a}. For electron spins in MOS QDs, the variations on the g-factor are typically $\Delta g/g\approx 10^{-2}$~\cite{Ferdous2018}. These variations could be partially mitigated by Stark shift but the strengths experimentally demonstrated ($\Delta g/g\approx 10^{-3}$~\cite{Huang2017}) remain too low to for full compensation. The level of variability in G-tensor components for electron spins is also dependent upon the orientation of the static magnetic field with respect to the silicon crystal axes but when the field is aligned along the [100] axis the variability is strongly suppressed~\cite{Tanttu2019}. Operating in the low magnetic field regime will also reduce qubit-to-qubit frequency variations. After applying these strategies, residual variability in qubit resonance frequencies could be managed by gradient ascent pulse engineering (GRAPE)~\cite{Khaneja2005}. 

 
Finally, variability associated with process changes will also need to be rapidly evaluated at scale to provide statistical evidence of improvement. High-throughput characterization techniques, e.g. based on low temperature ($<4$~K) wafer-scale probe stations or cryogenic multiplexers~\cite{Puddy2015, Pauka2020, Wuetz2020}, will need to be developed to correlate process-induced variability, and help identifying routes for its control and optimization. Furthermore, given the time-requirement to tune multiple gate voltages to bias levels where qubits can be operated, the field will strongly benefit from the development of computer-assisted auto-tuning routines for qubit initialization and parameter extraction~\cite{Baart2016, moon2020}. Once sources of variability are minimised, techniques to cope with residual variation will need to be developed, perhaps by constructing machine learning models that anticipate qubit performance from room-temperature diagnostic data. 

\subsection*{Modelling}

In order to speed up our understanding of the parameters that have an impact on qubit performance, microscopic modelling is necessary. The methods to model qubit devices are inspired by techniques used in the CMOS community to understand, for example, disorder and scattering. The main differences with respect to standard CMOS modelling are that, for qubits, modelling needs to be done in the one/few charge regime and rather than simulating electrical currents, the models need to address charge densities and wavefunctions.

From a methodological perspective, finite volume Poisson solvers can initially be used to compute the electrostatic potential in the active region of the device. From there, two methods can be used to compute the figures of merit such as electron filling, valley-splitting, tunnel coupling, g-factor, exchange coupling strength, etc. The first calculates the $N$ single-particle states in the electrostatic potential using either a multi-band $k\cdot p$ or a tight-binding (TB) model~\cite{Venitucci2019_2}. The second leverages existing tool suites and modified effective mass theory~\cite{Mohiyaddin2019}.

Up to now, these tools have mostly been used {\em a posteriori} to explain valley-splitting~\cite{Gamble2016, Ibberson2018}, EDSR for electron spins~\cite{Bourdet2018a, Bourdet2018b} and to model Rabi frequency~\cite{Venitucci2019}. Because more statistical data are now being generated, it might be possible to actually build a complete QCAD (Qubit Computer Aided Design) suite that goes from a microscopic description all the way to qubit array simulation.

\section*{\label{lowTCMOS} Challenges for cryo-CMOS design}

Designing electronic circuits at deep cryogenic temperatures poses some challenges that apply both to the quantum-classical interface and the classical layer. In the following subsections, we describe the impact of temperature on device parameters, power dissipation restrictions and communication latency. 

\subsection*{Temperature effects at the device level}

Conventional integrated circuit design uses established transistor compact models and passive circuit equivalents that enable predicting the circuit performance before manufacture. Those models needed to be redeveloped for cryogenic temperature operation. In a preliminary exploratory phase, individual technologies were studied at low temperatures which enabled establishing some important initial rules of thumb about transistor performance at cryogenic temperatures and allowed constructing preliminary models~\cite{Kamgar1982, Hanamura1986, Balestra1987, Broadbent1989, Balestra1994, Simoen2001, Yoshikawa2005a, Hong2008, Coskun2014, Homulle2019}. 

Bipolar technologies and old CMOS technologies above the 160~nm node are ruled out either because of freeze out of carriers below 4~K or because of the non-linear transistor behaviour, commonly referred to as the \lq\lq kink effect\rq\rq, which occurs at $V_\text{ds}\geq 1.2$~V~\cite{Homulle2019}. However, in general, modern CMOS technologies, both bulk silicon and fully-depleted silicon-on-insulator devices, operate at deep cryogenic temperatures although with modifications that present a weak temperature dependence below 4~K. Firstly, the threshold voltage ($V_\text{th}$) increases because of bandgap widening, carrier density scaling, and incomplete ionization with typical enhancements for n-type devices of 0.1-0.2 V~\cite{Beckers2019, Yang2020}. Such shifts may compromise technologies with low supply voltage ($V_\text{dd}$), pointing towards low $V_\text{th}$ or back-gated silicon-insulator-technologies as optimal choices for deep cryogenic design. The detrimental effect of the increase in $V_\text{th}$ on the on-state current, $I_\text{on}$, is partially compensated by an increase in mobility because of the reduced phonon population. Furthermore, the subthreshold swing in MOSFETs decreases because of the reduced thermionic transport down to the level of ~10 mV/dec where it saturates to a value proportional to the extent of the conduction-band tail associated with shallow defect states~\cite{Bohuslavskyi2019, Beckers2019a}. Although this represents a substantial reduction with respect to room temperature, it is far from the Boltzmann limit of 0.8~mV/dec and 20~$\mu$V/dec at 4~K and 100~mK, respectively. Furthermore, figures of merit such as on-off current ratio, $I_\text{on}/I_\text{off}$, and $g_\text{m}/I_\text{D}$, significantly improve at 4~K, aspects that are expected to enhance the performance of both digital and analogue circuits at cryogenic temperatures. In terms of passive components, quality factors improve at low temperatures~\cite{Patra2020b}. However, there are also disadvantages: Impedance mismatch deteriorates and $1/f$ noise becomes more prevalent due to the reduction of the thermal noise with respect to room temperature which creates different requirements when managing noise~\cite{Patra2018,Incandela2018, Hart2020}.

These initial findings have enabled moving to the next phase of cryogenic IC design in which heuristic knowledge and advanced compact models~\cite{Incandela2018, Galy2018, Beckers2019a, Beckers2019b} can be used to design analog and digital circuits to meet high-level specifications. In what follows, the field needs to move on to mass-scale characterisation of transistors and circuits to generate established cryogenic compact models and electronic computer-aided design (ECAD) tools for deep cryogenic temperatures.

\subsection*{Power consumption and communication latency}

Active QEC protocols require fast feedback between measurement and control. Utilizing a classical processing unit at room temperature to process the readout outputs and determine the gate sequences to correct for errors can be problematic. Sitting approximately 1.5~m away from the quantum layer, the distance imposes a minimum latency time of 30~ns which becomes comparable, for example, to two-qubit gate times mediated by the exchange interaction~\cite{He2019}. A distributed computer architecture will ultimately limit the bandwidth and pose synchronization challenges. Cryo-electronic circuits in close proximity with the qubit layer are then desirable to reduce the impact of latency on the efficiency of QEC protocols. However, dynamic operation at cryogenic temperatures puts some tight restrictions on the power budget and presents a concern of up to what level co-integration at millikelvin temperatures, where qubits operate best, may be possible. At 4~K the available cooling power is a few watts whereas at 100~mK is typically below 1~mW. Considering that transistors will have to be (dis)charged at radio or microwave frequencies,  dynamic power dissipation is a concern. But several solutions may be put in place to address this challenge:

First of all, recent results indicate that the operation of silicon spin qubits may be performed within error correction thresholds at elevated temperatures (1.1-1.45~K) bridging the gap between the quantum and cryo-electronics by enabling a higher cooling power budget~\cite{Petit2018, Yang2020a, Petit2020}. It must be noted that, so far, operation at higher temperature has been achieved at the cost of reduced fidelity. QEC correction will then require a larger number of physical qubits and associated classical electronics, increasing the overall power consumption. The optimal trade-off between fidelity and cooling power will need to be studied and determined. Further research directed to overcome the deterioration of qubit performance as temperature is increased will benefit the field, for example by using readout methods based on Pauli spin blockade which can provide high fidelity at elevated temperatures~\cite{Urdampilleta2019} or by minimising the increase of charge noise as temperature is increased~\cite{Petit2018}.

Secondly, a new branch of IC design should emerge with the objective to deliver dynamic performance with ultra-low power consumption to meet the demanding specifications for low temperature operation. Since dynamic power dissipation increases with the square of the voltage, the strategy could entail exploring technologies and designs that can operate at ultra-low supply voltages of just a few hundred millivolts. Given the substantial decrease in sub-threshold swing at deep cryogenic temperatures this is a realistic proposition. Here, technologies with $V_\text{th}$ tuning capabilities like back-gated SOI or even new technology nodes with $V_\text{th}$ designed for optimal low-power cryogenic operation may provide a solution. We restrict our discussion of power consumption to dynamical rather than static power consumption, the latter being dominated by leakage currents through the channel, which are substantially reduced by two to three orders of magnitude at deep cryogenic temperatures~\cite{Beckers2019b}.

Thirdly, since dissipation occurs on resistive elements, superconducting interconnects may be used to reduce the impact of dynamic losses, increase signal transmission and reduce the phononic heat flow from the cryogenic electronics to the qubits. However, power is still required to charge transistor gates and cable capacitances which, if not operated adiabatically, will eventually be dissipated. Low-power superconducting electronics such as RSFQ~\cite{Likharev1991} or nTron circuits~\cite{McCaughan2014} could also be used instead of CMOS but are yet less evolved. 

\section*{\label{QCint}Challenges at the quantum-classical interface}

As opposed to classical circuits, quantum computers need to have every logic gate individually controlled by external inputs. Furthermore to control and readout spin qubits, high-frequency analog signals are needed. In the following subsections, we describe these aspects that pertain the quantum-classical interface. 

\subsection*{\label{IO}Signal routing (I/O management)}

Brute force approaches by wiring each qubit and exchange gate electrodes to room temperature electronics do not scale well as they require macroscopic electrical wiring and extensive heat load management. Efficiently delivering control and readout signals to increasingly more complex quantum circuits, while reducing the number of room temperature inputs per qubit, is a key challenge in developing a large-scale universal quantum computer~\cite{Franke2019, reilly2019}.

One solution is to explore shared-control approaches in which QDs, as well as tunnel barriers, are controlled by common gates, mimicking the structure of CCD sensors. The approach relies on a high level of uniformity at the quantum layer, for example, by loading a single electron on individual QDs with a single common voltage. It has been recently shown that the standard deviation on the voltage required to load the first measurable electron on 40~nm silicon CMOS technology approaches this requirement~\cite{Yang2020} but further improvements will need to be made. Furthermore, variability in the tunnel coupling will be important in shared control schemes requiring spin shuttling and would need to be engineered to be controllable within an order of magnitude across the array~\cite{Li2018}. High quality material stack and fabrication techniques will be needed to achieve this level of uniformity and this is were advanced CMOS fabrication could make a difference.

While advances are made on reducing variability, parallel efforts should focus on developing addressing methods that deliver independent signals to each gate with a smaller number of resources, for example by using row-column addressing methods as in dynamic random access memory (DRAM) and CMOS image sensors. Qubit cell matrices with row-column addressing could provide independent control over $N$ qubits with O($\sqrt{N}$) room temperature resources by sacrificing simultaneous operation~\cite{Veldhorst2017}. Memory functionality will need to be added in the form of floating capacitors to retain voltages at the relevant gates for a period much longer than the coherence time of the qubits~\cite{Pauka2021,xu2020, hasler2021}. To extend the retention time and minimise unwanted voltage drifts at the gate, thicker gate oxides and low operation gate voltages will be beneficial. Efficient readout of such arrays could be achieved by using mixed high-frequency readout methods combining FDMA and TDMA~\cite{Ruffino2021a}. 

\subsection*{\label{cryoCMOS}Control and readout electronics}

A common feature for control and readout electronics is the necessity to generate IQ-modulated radio or microwave signals while additionally, for readout, reflected signals need to be demodulated. A generic architecture amenable to integration is one of a radio transceiver, with the difference that it must be designed to operate at cryogenic temperatures. Figure~\ref{Fig2}b shows such an architecture~\cite{Charbon2016}, where one can recognize the readout path, represented by multiplexers, amplifiers, demodulators, and ADCs. Typically, the voltage of the devices to be read is just a few tens of microvolts and, when using reflectometry techniques for sensing, the reflected amplitudes to be measured may even be smaller. Therefore, there is a need for amplifying the signals before demodulating them. Traditionally, this task is performed by a cryogenic high-electron-mobility low-noise amplifier (LNA), with low noise equivalent temperature ($T_\text{N}$) -- this figure determines the noise level of the measurement and is typically of the order of a few Kelvin. Furthermore, it requires an amplification of the order of 40~dB or more to minimise the impact of subsequent amplifying stages placed at room temperature. Circulators are typically placed in between the QPU and the LNA to minimise interference between forward and backward travelling waves (not shown). Readout can be multiplexed in time (TDMA) and/or in frequency (FDMA). In the latter case, the bandwidth of the LNA becomes an important factor due to the need to spectrally pack several readout tones in the same channel. Thus, typically bandwidths of the order of a gigahertz or higher are desirable. Although LNAs are typically manufactured using InP, SiGe-based amplifiers with $T_\text{N}=2$~K have been demonstrated and could be integrated with CMOS or SiGe BiCMOS~\cite{Weinreb2007}. Recently, several of these elements have been integrated on a chip to provide a compact alternative to distributed readout electronics for high-frequency gate-based readout: (i) an integrated CMOS transceiver addressing the 6-to-8 GHz band including, on a single chip operating at 4~K, RF amplifiers, I/Q downconversion mixers, and baseband amplifiers and filters~\cite{Prabowo2021} and (ii) a fully integrated 5-to-6.5 GHz I/Q receiver operating at 3.5~K, also incorporating a low-noise MW amplifier as well as readout signal generation~\cite{Ruffino2021}.

To reduce further $T_\text{N}$, quantum-limited amplifiers such as the Josephson parameter amplifier (JPA) could be used in conjunction with dispersive readout sensors~\cite{Schaal2020}, an approach routinely used for superconducting qubits~\cite{Arute2019}. The JPA in phase-preserving mode enables reducing the readout time by an order of magnitude with respect to conventional cryogenic amplifiers and in phase-sensitive mode could enable going beyond the quantum-limit using quadrature squeezing. For large amplification bandwidth necessary for FDMA, travelling wave amplifiers (TWPA) may be used. And for full-integration, parametric amplifiers based on the dissipationless quantum capacitance of silicon QDs could be explored~\cite{Esterli2019}.  

The control path in Fig.~\ref{Fig2}b shows DACs, IQ modulators and rf amplifiers aimed to assist controlling the qubits, potentially through multiplexing. Spin control is achieved through local oscillators that are IQ modulated to create a series of carefully timed envelopes. In recent demonstrations using cryo-CMOS integrated circuits \cite{Bardin2019,Patra2020,Dijk2020}, operation has been based on a programmable frequency, spanning from 1 to 20~GHz, so as to allow for sufficient flexibility in the type and number of qubits that can be controlled. More recently, the digital section of the control has grown substantially, to perform sophisticated modulation of $4\times32$ frequencies, which are upconverted only in the last rf stage~\cite{Patra2020}. This approach has been used to control electron spin qubits in SiGe with the same fidelity as with commercial instruments at room temperature and could deliver the desired performance at power consumption compatible with 4~K operation for several tens of qubits simultaneously (1~mW/qubit)~\cite{Xue2021}. Recently, a fully integrated cryo-CMOS chip combining both control MW signals in the 11-to-17 GHz range and readout on the 200-to-600~MHz range based for rf-SETs has been demonstrated at 4~K~\cite{Park2021}. 

More generally, to link the impact of the control and readout electronics on qubit fidelity, a simulation framework SPINE (SPIN Emulator)~\cite{Dijk2019} has been developed. SPINE enables determining the minimum set of specifications for the control and readout electronics in order not to become the bottleneck in QPU performance. It does so by linking a qubit's time evolution with the time varying signals generated by the classical electronics.

\section*{\label{Classical}Challenges at the classical layer}

After the analog readout is performed, digital decoding of the qubit measurements and the formulation of a response to control them with an appropriate set of signals needs to be put in place. For this, the classical logic will need to operate significantly faster than qubits to read, identify the error, and produce the response with high fidelity. Even if the operation timescales for silicon qubits could be pushed down to 10 or 100~ns, this could still be managed with current classical processors with clock rates at 3~GHz.  However, in terms of classical processor performance, error decoding at scale may pose a challenge. Assuming QEC cycles~\cite{Fowler2012} in the submicrosecond regime, reading a million qubits and determining whether errors have occurred will require processing $>1$~Tbits/s. The most suitable processor architectures to handle the data efficiently while minimizing power consumption will need to be investigated.  

Digital circuits are much less problematic than analog, due to the large noise margins. The first circuits have been based on reconfigurable architectures, in particular FPGAs~\cite{Lamb2016,Homulle2017,Homulle2019}. FPGAs could operate normally in deep-cryogenic temperatures and all the functions could be activated. Independently, QEC algorithms have been designed and implemented in room temperature FPGAs~\cite{Fu2018}. The current trend is to combine these advances into FPGAs operated at cryogenic temperatures improving the feedback speed and reducing latency in QEC or enabling surface code decoding through machine learning techniques~\cite{Varsamopoulos2020}. FPGAs will likely be used to enable fast back and forth communication in QEC but also in hybrid quantum-classical algorithms like the Variational Quantum Eigensolvers (VQE)~\cite{Peruzzo2014} and quantum approximate optimization algorithms (QAOA)~\cite{farhi2014}.

Another aspect relevant to classical logic performance at low temperature is that reduced leakage currents can make DRAMs essentially static, thereby enabling significant reduction in real estate utilization or alternatively in the increase of memory available to the controller. A considerable increase of memory could enable more sophisticated waveforms to counter the effects of simultaneous control of an increasing number of qubits and to enable true scalability. 

\section*{\label{architecture}Challenges at the architecture level}

To build a QCS, the quantum and classical layers and their interface need to be assembled in a functional manner. This raises the question of what the best system configuration is and what temperature the different function blocks should be placed at. These layers can be almost independently optimised. So far, the approach has been bottom-up, thinking of the optimal quantum layer and building all the way up.

One of the big advantages of silicon in terms of scaling, its small qubit footprint, also poses some challenges in terms of system design: The I/O problem, and the location of the classical electronics with respect to the quantum layer. All considerations apart from that of power consumption indicate that monolithic integration would be optimal. However, if the classical electronics is co-located in the plane with the quantum layer, the 2D array necessary for QEC would be broken. With these requirements in mind, several architectures have been proposed with varying levels of integration going from full 3D integration to 2D modular designs to account for limitations, leverage know-how and validate elementary designs (Fig. 4). 

3D Integration: The current variability of the voltages used to confine the charges and to tune the coupling between adjacent QDs, suggests individual gate addressing. To manage the large gate overhead researchers have proposed to co-locate floating memory units with embedded control transistors (similar to 1T-1C DRAM modules) to minimise the number of I/O connections~\cite{Veldhorst2017}. These memory units enable individual biases for each QD and exchange gate in a row/column addressed matrix but require of periodic voltage refreshing. Though indeed this architecture solves the challenge of managing local variability and the I/O problem, this proposal relies on aggressive technological assumptions: short gate pitch, 3D integration of quantum and classical electronics, and the management of crosstalk between signal in the gigahertz regime, aspects that have not yet been routinely demonstrated by the semiconductor industry.

An alternative monolithic 3D integrated architecture has been proposed with the focus on integrating readout on chip and providing a simpler way of initializing the qubits using two-layers: the bottom one for electrometre and reservoir definition, and the top one to encode qubits~\cite{Vinet2018}. One advantage of the proposal is that enables implementing the surface code in a $2\times2$ QD sublattice since every qubit has a dedicated rf sensor capable of spin readout, whereas the aforementioned design requires a $2\times3$ sublattice~\cite{Veldhorst2017} due to the necessity to do projective Pauli spin blockade readout (Fig.~\ref{Fig3}b).

2D Integration: Subsequent proposals simplified the architecture by moving to planar designs. The floating gate arrangement was replaced by a 2D monolithic design including three gate layers: row, column and diagonal, to control the QDs and the horizontal and vertical exchange, respectively~\cite{Li2018}. Readout and control electronics is displaced to the periphery by making use of dispersive readout and multiplexed microwave signals. Although compatible with current technology, the architecture relies on shared control that requires a level of uniformity still to be demonstrated.

2D Modular: Recently, researchers proposed modular 2D sparse geometries to relax the fabrication and variability constraints~\cite{Vandersypen2017, Boter2019, Petit2020}. The concept offers flexibility to accommodate the gate layouts to individually tune QDs and exchange interactions in local registers and introduces the concept of long-distance coupling between spatially separated registers. Separated qubit registers could alleviate wiring problems, reduce crosstalk and free up space to co-locate classical electronics modules. Several methods for long-distance coherent coupling over distances varying from microns to millimetres exist but the most widely studied rely on charge or spin shuttling~\cite{Flentje2017,Hermelin2011} or microwave-photon-mediated spin-spin interactions~\cite{Borjans2020}. For shuttling, two mechanisms exist: (i) Surface acoustic waves (SAW)~\cite{Bertrand2016} and (ii) CCD-inspired tunnelling between adjacent QDs~\cite{Fujita2017, Mills2019, Mortemousque2021}. SAW-based spin transfer is most effective in piezoelectric materials and has been used to displace an electron spin between two AlGaAs/GaAs-based QDs separated by 6~$\mu$m with up to 90~\% fidelity~\cite{Jadot2021}. The feasibility of this mechanism in silicon remains to be investigated but it will possibly require the use of additional piezoelectric materials. On the other hand, CCD-inspired shuttling has been demonstrated in silicon: individual electrons have been transferred over micrometer distances~\cite{Mills2019} and coherent spin shuttling has been achieved between two adjacent QDs with an average fidelity of 99.4\%~\cite{Yoneda2020}. Demonstrating coherent spin transfer in a QD array spanning micrometres length scales with high fidelity remains to be demonstrated. It would be necessary to minimise the local g-factor differences between QDs with the strategies described above. Cavity-mediated interactions via real or virtual microwave photons are not yet at the threshold for fault tolerance due to the difficulty to achieve coherent coupling rates between a single spin and a photon largely exceeding the spin and photon decay rates. However, results on large coherent coupling rates in industry-fabricated double QDs~\cite{Ibberson2021} and the progress on developing high-impedance resonators~\cite{Zheng2019} may lead to fault-tolerant fidelities in the near future.

\section*{\label{sec:Large}Outlook}

We have examined how scaling silicon-based quantum computing could benefit from using CMOS fabrication lines and have drawn attention to the system-level design of a quantum computer: a holistic perspective that includes the quantum layer, the quantum-classical interface and the classical processing unit. Silicon offers the enticing advantage that all these different layers could, in principle, be manufactured using CMOS processes opening opportunities for compact system integration and low-cost manufacturing. Small silicon-based quantum processors may be readily manufacturable using CMOS technology; for large-scale fault-tolerant processors further development beyond the current capabilities of VLSI technology will be required but we foresee no fundamental roadblock. We have highlighted here the key engineering challenges and provided directions for how to address them.

The transition from fabrication at industrial-grade research and technology organizations to standard silicon foundries provides an opportunity for all of the different layers of a QCS. In particular, establishing a global Multi-Project Wafer prototyping service for silicon quantum circuits, with a larger flexibility in the violation of design rules, could have profound influence in the development of silicon-based quantum computing. It could lead to standardization, increased fabrication throughput, and improved accessibility to quantum devices and circuits. It could also reduce the timescales to the final goal: creating a large-scale fault-tolerant quantum computer. 

The realization of large-scale quantum computer will require a synergy between researchers with expertise in the control of elementary quantum systems and specialists in systems engineering from the semiconductor industry. Of equal value will be the promotion of specialised degrees that explore the boundaries between quantum physics and engineering, in order to train new professionals with simultaneous expertise in Pauli matrices and Verilog-A. We hope this Review will help stimulate the curiosity of the required range of researchers — from specialists in the CMOS community to new graduate students — who can join the exciting endeavour of building a large-scale silicon-based quantum computer.

\begin{acknowledgments}
We thank David J. Reilly, Andrew J. Ferguson, Arne Laucht, Andre Saraiva and Lisa A. Ibberson for providing useful comments. This research has received funding from the European Union's Horizon 2020 research and innovation programme under grant agreements No. 951852, 688539 and 810504. M.F.G.Z. acknowledges support from the Royal Society and the Winton Programme for the Physics of Sustainability. S.D.F., T.M., and M.V. acknowledge support and from the Agence Nationale de la Recherche, through the CMOSQSPIN project (ANR-17-CE24-0009). A.S.D. acknowledges support from an Australian Research Council Laureate Fellowship (FL190100167).

\end{acknowledgments}

\section*{Author contributions}
All authors contributed to the writing of the manuscript.

\section*{Competing Interest}
M.F.G.Z is employed by Quantum Motion Technologies a start-up focusing on building a silicon-based quantum computer. 

E. C. holds the position of Chief Scientific Officer of Fastree3D, a company making LiDARs for the automotive market, and he is co-founder of Pi Imaging Technology, maker of sensor for microscopy. Both companies have not been involved with the paper drafting.

M.V., S.D.F., T.M. and A.S.D. declare no financial competing interest. 


\clearpage

\section*{Figure Titles/Captions}

\begin{figure*}
	\centering
		\includegraphics[width=1.00\textwidth]{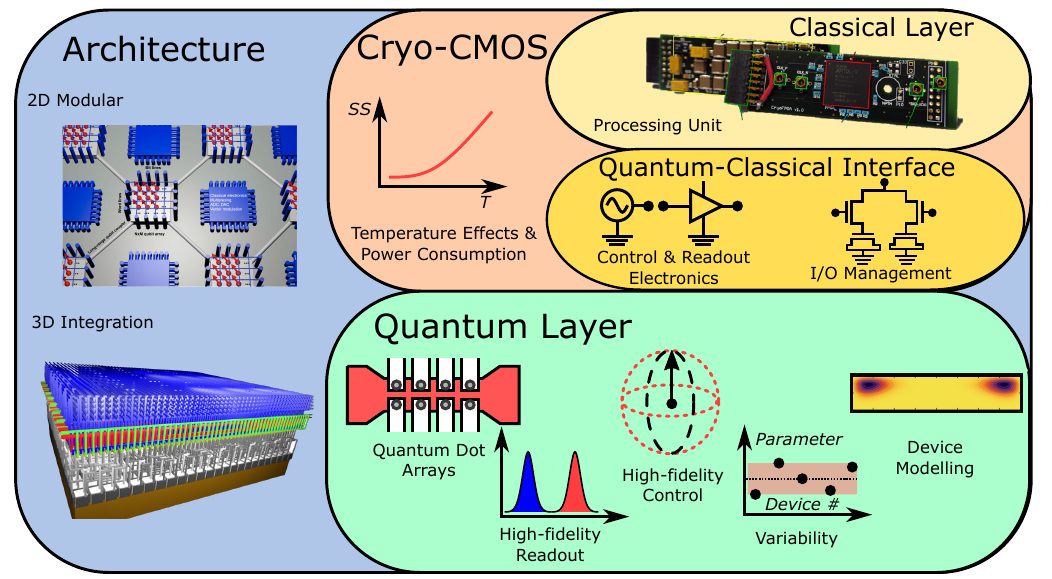}
	\caption{Challenges. Summary of the different challenges to scale silicon-based quantum computers using CMOS technology divided into the Quantum Layer, the Quantum-Classical Interface, the Classical Layer and the Architecture. The 2D modular and 3D integrated schematic architectures are from ref.~\cite{Vandersypen2017} and ref.~\cite{Veldhorst2017}, respectively. The photograph of the FPGA, indicated in red, in the section Classical Layer is from ref.~\cite{Homulle2017}. The simulation exemplifying device modelling is from ref.~\cite{Ibberson2018}.}
	\label{fig:Figure4}
\end{figure*}

\end{document}